\documentstyle[aps,twocolumn,epsfig]{revtex}

\begin{document}

\title{Noise in neurons is message-dependent \\
\tiny{(neuronal reliability/temporal coding/information theory/error-correcting) }}
\author{Guillermo A. Cecchi$^{\dagger}$, Mariano 
Sigman$^{\dagger}$ $^{\ddagger}$, Jos\'{e}-Manuel Alonso$^{\ddagger}$, Luis 
Mart\'{\i}nez$^{\ddagger}$, Dante R. Chialvo$^{\dagger}$, Marcelo O. 
Magnasco$^{\dagger}$} \address{{\sl $^{\dagger}$Center for 
Studies in Physics and Biology,  and $^{\ddagger}$Laboratory of 
Neurobiology,  The Rockefeller University, \\
 1230 York Avenue, New York NY10021, USA}} 

\maketitle

\begin{abstract}

Neuronal responses are conspicuously variable. We focus on one particular
aspect of that variability: the precision of action potential timing. We show
that for common models of noisy spike generation, elementary considerations
imply that such variability is a function of the input, and can be made
arbitrarily large or small by a suitable choice of inputs. Our considerations
are expected to extend to virtually any mechanism of spike generation, and we
illustrate them with data from the visual pathway. Thus, a simplification
usually made in the application of information theory to neural processing is
violated: noise {\sl is not independent of the message}. However, we also show
the existence of {\sl error-correcting} topologies, which can achieve better
timing reliability than their components.  

\end{abstract} 
\pacs{PACS:05.40.+j,87.19.La }

\section{Introduction}

Brains represent signals by sequences of identical action  potentials or
spikes \cite{adrian}.  Upon presentation of a stimulus, a given neuron will
produce a certain pattern of spikes; upon repeating the stimulus, the pattern
may repeat spike for spike in a  highly reliable fashion, or may be similar
only "rate-wise", or some spikes may  be repeatable and others not. If
individual spikes can be recognized and tracked across trials, then their
timing accuracy can be ascertained unambiguously; this is not always the
case. The existence of repeatable spike patterns and the reliability of their
timing changes not only from neuron to neuron, but even for the same neuron
under varying circumstances. Bryant and Segundo \cite{BRYANT} first noticed
(in the mollusk Aplysia) that spike timing accuracy depends on the particulars
of the input driving the neuron. This intriguing property has received renewed
attention, notably among those searching for experimental evidence supporting
different theories of neural coding \cite{SEJNOWSKI,BIALEK}. In a study of 
the response of pyramidal neurons to injected current \cite{SEJNOWSKI} the
temporal pattern of firing was found to be unreliable when the injected
current was constant, but highly reliable when the input current was ``noisy''
and contained high frequency components. This study showed explicitly the
difference between the irregularity of the spike pattern (which was irregular
in both cases) as opposed to  the {\em reliability} or accuracy of spike
timing, and also highlighted the connection that ``natural''  stimuli are
noisy and contain sharp transitions. Similar results have been obtained from {\sl in vivo} recordings of the H1 neuron in the visual system of the fly
\cite{BIALEK}: constant stimuli produced unreliable or  irreproducible spike
firing patterns, but noisy input signals deemed to be more ``natural'' and
representative of a fly in flight yielded much more reproducible spike
patterns, in terms of both timing and counting precision. Although some
aspects of the last study have been challenged recently in \cite{WARZECHA},
the phenomenon of different reliabilities for varying inputs was reaffirmed:
see e.g. Fig. 1 in \cite{WARZECHA}. Thus one concludes that similar
observations of the timing precision of spikes can be made in very
different types of neurons under vastly different conditions, and that this
must be a very universal, almost basic characteristic of neuronal function. We
shall now show how these experimental observations follow from rather general
and elementary considerations, and discuss the possible implications for brain function and architecture.

\section{Stochastic spiking models}

The essence of this phenomenon can be demonstrated easily in toy models. A
simple example is the leaky integrate-and-fire (I-F) model \cite{KNIGHT}, which
assumes the neuron is a (leaky) capacitor driven by a current which simulates
the actual synaptic inputs. We add membrane fluctuations representing several
internal sources of noise (cluttering of ion channels, synaptic noise, etc.) to
obtain a system described by the following Langevin \cite{RISKEN} equation:  

\begin{equation} 
C \dot{V} = -g V + I(t) + \xi (t) \label{I} 
\end{equation}

\noindent
where $C$ is the cell membrane capacitance, $V$ the membrane potential, $gV$
the leakage term ($g$ is a conductance), $I(t)$ is the input current, and 
$\xi(t)$ is Gaussian noise, $ \langle \xi (t) \rangle = 0 $ with
autocorrelation $ \langle\xi (t) \xi (t') \rangle = \sigma \delta (t-t') $;
when the potential reaches the threshold $V_{0}$ an action potential is
generated and the system returns to the equilibrium potential $V_{e}$, here set
arbitrarily to zero.

Fig.1 shows the results of a numerical simulation of  Eq.~(\ref{I}) in response
to two different signals \cite{simula}, in both cases in the presence of
identical ``internal'' noise.  Following refs.
\cite{SEJNOWSKI,BIALEK,WARZECHA}, the left panels show a time independent input
and the neuronal responses to repeated stimulation, while the right panels
ensue from a ``noisy'' input and its responses. When $I(t)$ contains high
frequency components (right side panels) spikes are clustered more
tightly to the upward strokes of the input and overall there is less
trial-to-trial variability than in the time-independent case. The phenomenology
described in Fig. 1 captures the main feature that renewed interest in this
subject; i.e., that apparently more ``noisy'' (in the sense of ``severely
fluctuating'') inputs can generate less variable responses.

% figure 1 here
\begin{figure}[htbp] 
\centerline{\psfig{figure=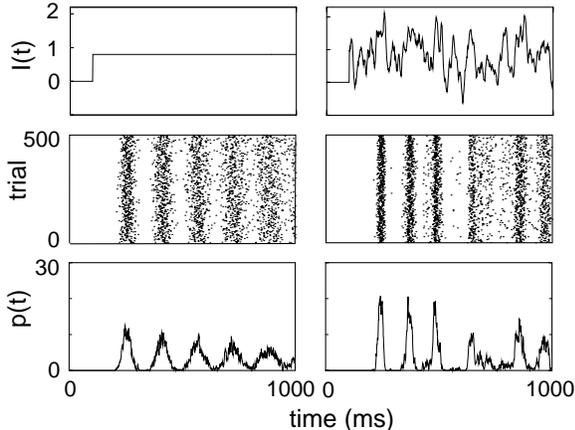,width=3.0truein}} 
\caption{\footnotesize {Rapidly fluctuating 
(panels on the right) inputs produce more reliable output spikes 
than  a time independent stimuli (panels on the left).
Panels in the top correspond to input stimulus - $I(t)$ in 
Eq. 1; middle ones to the 500 responses to the stimulus where the
occurrence of each spike is denoted with a dot; and bottom panels
depict the peri-stimulus time histogram describing the rate 
- $p(t)$ - at which spikes are generated in response to the stimulus
$I(t)$.} }
\end{figure}

By concentrating on these two extreme stimuli, these studies have somewhat
obscured an issue which is central to the variability phenomenon: the
relationship of the output to the timescale or timescales over which the input
itself varies. The constant stimulus can be thought to have no timescale, or at
best, just a timescale since onset; while in principle the noise process has
indefinitely fast timescales. Thus the two extreme cases under discussion have
timescales outside the range of interaction with physiological processes.

A simple dimensional argument shows that timing variability is dependent on the
characteristic timescale of the input. As a consequence of noise, the neuron's
membrane potential will have fluctuations of a characteristic size (or root mean square)
$\Delta V$. In order to translate this $\Delta V$ into a timing noise $\Delta
t$, we have to scale the potential noise by something having units of $ \Delta
t/ \Delta V $; the most natural such quantity at hand is the inverse of the
potential's time derivative. 

While the distribution of the derivative of the potential is a reasonable
measure of the degree of reliability of a particular input, one can see from a
geometrical argument that the derivative of the  potential at the particular
time in which the potential crosses the threshold  is a better estimator of
stimulus reliability. The uncertainty in firing time is essentially the time
during which the membrane potential is at a distance from the threshold smaller
than the value of the voltage noise, and so is scaled by the derivative at this
particular moment. Thus, {\it the faster the voltage approaches the threshold
(in the vecinity of the threshold), the more reliable the timing of the spike
will be.} The statement captures much of the phenomenology in
\cite{BRYANT,BIALEK}, and in our own data, as we will show.

% figure 2 here
\begin{figure}[htbp] 
\centerline{\psfig{figure=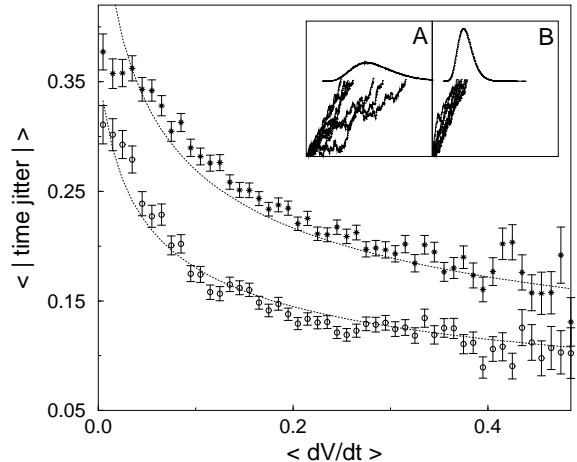,width=3.10truein}} 
\caption{\footnotesize{Computed mean spike temporal
jitter (and SEM) as a function of the estimated membrane 
potential's d$V$/dt. Estimates are plotted for two noise 
variances: $\sigma=0.01$ ($\circ$) and 
 $\sigma=0.02$ ($\star$). The expected inverse-law fit (with
exponent $\sim 0.35$) is depicted with dotted lines. The 
examples in the insets illustrate two density 
distributions of spiking times resulting from relatively 
slow (inset A) or relatively fast (inset B) d$V$/dt 
threshold crossings. For each case, below each histograms, 
a few typical trajectories are also plotted showing the membrane potential 
preceding the threshold crossing. }}
\end{figure}

We proceed now to apply this method of estimation to the situation shown in
Fig. 1. A rapidly fluctuating signal was constructed \cite{simula} and the
times at which spikes occurred were determined first for the deterministic
condition (i.e., $\sigma=0 $), and subsequently for several hundreds stochastic
realizations (i.e., with identical $I(t)$ but with different stochastic
realizations of the noise term in Eq. 1). At the same time the voltage
derivative preceding each spike was computed over the prior fifty time steps.
The fluctuating character of $I(t)$ (as illustrated in the top right panel of
Fig. 1) offers the opportunity to explore a wide range of d$V$/dt at threshold
crossings. To estimate the precision of spike timing we define an index of
``temporal jitter'' as follow: it is the absolute difference between the time
at which a spike is generated in the noise-free simulation and that of the
corresponding nearest spike for the stochastic realization, averaged over
realizations, and over all spikes which occurr within a d$V$/dt interval of
$0.1$. This quantity is plotted in Fig. 2. The data points in Fig. 2 are
scattered around the expected (dotted lines) inverse relationship between the
temporal jitter of the spike and the speed at which the voltage crossed the
threshold. The trajectories plotted in both insets  help to visualize the
geometrical argument already discussed: inputs that rise more rapidly to the
membrane potential threshold produce less variable spike timing.

The same basic phenomenon will affect any model of spike
generation, in which the noise in the voltage-like
variable is translated into jitter of its threshold crossings. To
provide for a specific example, we choose a widely used model of
excitable media with continuous dynamics, the FitzHugh-Nagumo model
(FHN) with additive stochastic forcing \cite{CHIALVO}, described by the
following Langevin system:

\begin{eqnarray} 
\dot{V} & = & V - V^{3}/3 - W + I(t) + \xi (t) \\
\dot{W} & = & \phi ( V + a - b W) 
\end{eqnarray}

The variable $V$ is the fast voltage-like variable, $W$ is the {\it slow}
recovery variable; $a$, $b$ and $\phi$ are constant parameters, and $\xi (t)$
is a zero-mean Gaussian white noise of intensity $D$. According to these
equations, after a spike the recovery produces an absolute refractory period
during which a second spike can not occur, followed by a longer relative
refractory period during which firing requires stronger perturbations.

% figure 3 here
\begin{figure}[htbp] 
\centerline{\psfig{figure=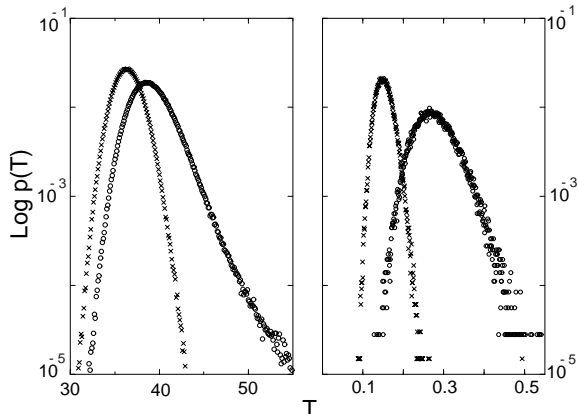,width=3.truein}}
%\vspace{.0in} 
\caption{\footnotesize{First passage time distributions for the  FHN
(left panel) and I-F (right panel) models, forced with constant input.  In both
panels, ($\times$) corresponds to high input and ($\circ$) to low.} } 
\end{figure}

As previously argued, it is biologically reasonable to assume that noise
affects the voltage-like variable. Thus, according to our argument, a higher
steady input will result in a faster approach to the threshold crossing and
therefore will reduce the probability of an untimely crossing. This is
confirmed in Fig. 3, where the distributions of periods are plotted for two
inputs amplitudes. Note the large change in variance (1.5 to 2.34, 56$\%$) for
comparably small change in the average period (36.4  to 39.1, 7$\%$) in going
from high to low input, also note that the low input distribution has a
supra-exponential tail. For comparison, the distribution of inter-spike
intervals corresponding to the I-F model with constant input is shown. In both
cases, the distribution for high input tends to a Gaussian (parabola on
semi-log plot), while the low input has an exponential tail (linear on semi-log
plot) as expected for Poisson-like statistics. It is interesting to notice that
even though this phenomenon will affect more strongly systems with
zero-frequency quiescent-to-firing bifurcations, given the longer time to
integrate fluctuations at low input (i.e. the system can evolve near a
homoclinic orbit), the example presented here shows that it also affects
non-zero-frequency transitions like the Hopf bifurcation present in the FHN
model.

\section{Experimental results}

Thus, in neural data, one should expect to find that the output noise
generically depends on the structure of the input. Experiments discussed next
demonstrate the relevance of this observation for brain function. We recall
that an assumption usually made in the application of information theory to
neurons is that the noise introduced by the communication channel is
independent of the message being transmitted, and now explore to what extent
this assumption is valid in real neural data. The result is presented in Fig.
4, where different visual stimuli (i.e., the four messages) were presented to
an anesthetized cat while electrophysiological recordings were made in the
lateral geniculate nucleus (LGN), which is the second stage of processing in
the visual pathway (see \cite{cat1} for methods).

The raster plot in Panel A shows the response to a moving bar with high
contrast and high speed, and Panel B shows the responses to a low contrast bar
moving at low  speed, both results are shown in a window of 200 msec centered
at the peak of the ON response. The responses in Panel A clearly display a
higher temporal precision than those for the condition in Panel B. Two further
conditions were also presented, for high contrast and low speed, and for low
contrast and high speed. In the inset of Fig. 4 we quantify the noise for the
four messages as the ``jitter'' of the response onset, which we define as the
average of the absolute value of the position with respect to the mean of the
first spike in each trial. The messages include two conditions for the velocity
and two for the contrast. For each velocity  condition, the high contrast is
more reliable than the low one: 8.4 msec vs. 22.1 msec, and 16.8 msec vs. 23.8
msec. Note however that the high contrast/high speed and low contrast/low speed
stimuli, although they have a very similar spike response (4.17 vs. 3.86) have
the largest difference in jitter (8.4 msec vs. 23.8 msec). Alternatively, we
can measure the entropy difference between the corresponding first spike
probability distributions (1 msec bins), which varies from 2.0 bits for low
contrast/high speed vs. high contrast/high speed, to 0.6 bits for high
contrast/low speed vs. high contrast/high speed. Similar results showing
contrast-dependent temporal precision in LGN have been reported previously in
\cite{REICH}, where it is also proposed that the variability is due to
intrinsic noise. We also show that messages with indistinguishable mean rate
responses can show large differences in their variability, in agreement with the hypothesis of the threshold crossing derivative as the relevant parameter, which can be independent of the firing rate for particular inputs.

% figure 5 here
\begin{figure}[htbp]   
\centerline{\psfig{figure=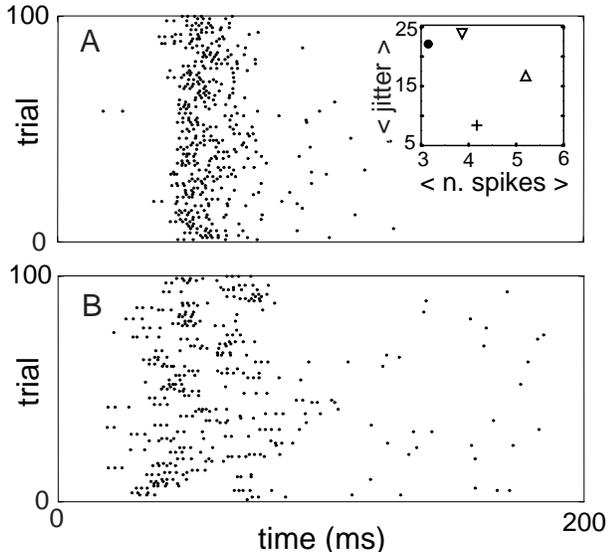,width=3.15truein}} 
%\vspace{.1in}  
\caption{\footnotesize{Experimental example showing that  different
``messages'' elicits different noise (i.e., spike  temporal jitter). Results
gathered from recordings in cat's LGN  where neuronal action potentials are
recorded in response to the  presentation of a moving bar with  different
contrast and speed. The inset in Panel A shows, for the conditions tested,
the  temporal jitter versus average number of spikes (collected during the
window of 200 ms); ($\bullet$) corresponds to low contrast/high speed, (+) to
high contrast/high speed,  ($\bigtriangledown$) to low contrast/low speed, and
($\bigtriangleup$) to high contrast/low speed. Notice the data from moving bars
of different contrast originating very different jitter. The raw data for two
of the experimental conditions are  presented in the raster plots of Panels A
and B. Responses  obtained with a moving bar of low contrast moving at low
speed are plotted in Panel B. The (less variable) responses obtained with a
moving bar of high contrast moving at high speed are  depicted in Panel A.}} 
\end{figure}

\section{Network architecture and reliability}

The observations discussed in the previous section are restricted to the
behavior of individual neurons. How can they be extended to networks of spiking
elements? In principle, we can assume that input-dependent noise will be
present in early stages of processing, given the relative simplicity of the
neural architecture. The relevance of this phenomenon in higher order areas
remains to be explored; nevertheless, we may hypothesize that unless dedicated
architectures are implemented to eliminate it, significant input-dependent
noise will be ubiquitous in neural processing.

Indeed, a recurring question in brain theory and computer theory is whether
there are reliable ways to compute which use unreliable components. We shall
now show explicitly the existence of a network topology for which the overall
time reliability can be made arbitrarily better than that found in its
individual neurons. 

Assume a single input source $I(t)$, which ``fans out'' to $N$ noisy neurons
$X_i$; these in turn ``fan in'', or input into a single (also noisy) neuron
$Y$. Let's imagine an $I(t)$ constructed so that the $X_i$ fire once each at
times $t_i$. If the size of the connections $X\to Y$ is set so that $Y$ fires
after seeing half of the expected number of spikes coming from the $X$s, then
clearly neuron $Y$ will be fire near the {\em median} time of the spike times
$t_i$. The median, as a descriptor of the $t_i$, has a number of  exceedingly
nice properties in this context: (a) its variance decreases as $1/\sqrt{N}$, 
and thus the timing accuracy of $Y$ can be made better than the individual
$X$s simply by choosing an appropriately large $N$; (b) the median, unlike the
mean, is exceedingly robust against outliars and heavy-tailed noise; (c) the
median does not require that it see the entire set $t_i$ but only half of it,
and (d) the median is expected to lie at the time of highest concentration of
the $t_i$, thus maximizing the timing accuracy of $Y$. This can be fully
apreciated in Fig. 5. Thus, we see that different topologies will propagate
spike timing accuracy in different fashions, and that one probably should
expect architectural correlates in place in the brain. 

To gain further insight into this issue and to study the implications of our
observations on the synchronization of a neuronal network, we now concentrate
on random but fixed time delays across pathways, rather than on individual
timing accuracy. We modify the previous model: all neurons of the middle layer
are now noise-free and in principle identical, and so are their spike trains;
the neurons are integrate-and-fire units as in Eq. 1, with the further addition
of a refractory period. We next model a dispersion of the transmission delay
times by rigidly shifting the spike train of each individual neuron by a random
value uniformly distributed between $0$ and $\delta$. The total charge that the
integrating neuron receives does not depend on this shift, but when all neurons
are synchronized, the derivative of the charge will be high and then we expect
more reliable responses. Fig. 6 shows how the timing reliability (open
circles), measured by the method of spike metrics \cite{VICTOR}, decreases with
the decorrelation of the input layer. Interestingly, when the derivatives are
high the response is more reliable, but by the same token, the current supplied
during the refractory period will be lost. This means that smooth distributions
will result in an increased effective charge and thus the rate of response of
the integrating neuron is bigger when the input layer is decorrelated, as shown
in Fig. 6 (open squares) \footnote{There is a second regime (not shown), if
the decorrelation is large compared to the decay time of the neuron, in
which rate decreases with increasing $ \delta $ due to current loses.}.

% figure 5 here
\begin{figure}[htbp]
\centerline{\psfig{figure=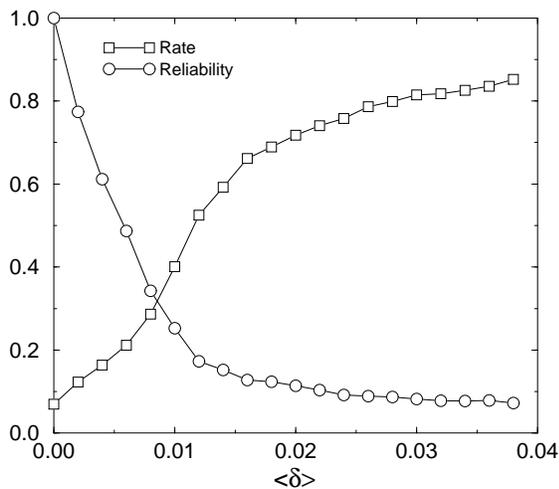,width=3.4truein}}
\caption{\footnotesize{
Simple model to study the effects of synchronization on the reliability of the
neuronal responses. 75 Neurons synapse on a unique neuron modeled as a leaky
integrate-and-fire as in Eq. 1, with the addition of a refractory period;
g=0.005/s, $\sigma$=0.07, refractory period = 10 $ms$.
Mean Rate and Reliability for a neuron integrating spikes over an input layer
of 75 neurons. A pattern of spikes was randomly generated (3.75 Hz) and each
neuron of the input layer feeds the same pattern of spikes with a delay
$\delta$. The value of the other parameters are as in Fig 5. When the input
layer is synchronized, $\dot {V}$ of the integrating neuron is big and the
response is very reliable. In this case, part of the charge from the input
layer is lost because there is no summation during the refractory period. When
$\delta$ increases, the reliability decreases but the mean firing rate is
increased. This suggests that a function of synchronization, which is known to
occur in many different neuronal networks, might be to increase the reliability
on the system at the expense of the intensity of the signal.
}}
\end{figure}

\begin{figure}[htbp]   
\centerline{\psfig{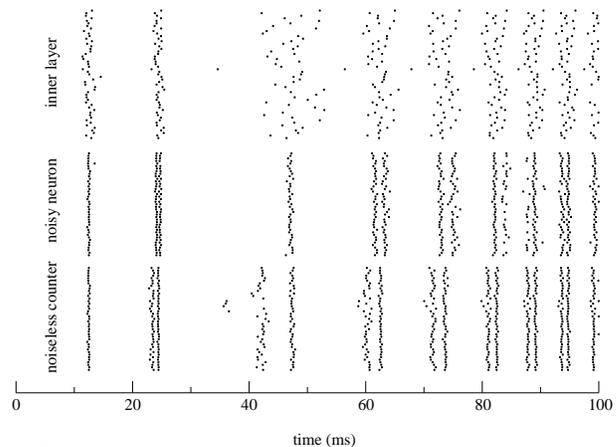}} 
%\vspace{.1in}  
\caption{\footnotesize{Reliability of a fan-out/fan-in structure. 
An input $I(t)$ is fanned out in parallel to $N$ identical neurons; one extra
(identical) neuron fans-in: it receives input from all $N$, with the weights 
adjusted so as to fire once per $N/2$ input spikes. In this example, $N=51$
and $I(t)$ is a constant plus an Ornstein-Uhlenbeck process. Top panel:
rasters for {\em one} instance and all 51 neurons, to illustrate the 
reliabilty of the middle layer. Middle panel: raster for the output neuron,
40 instances. We emphasize that the output neuron has {\em identical} parameters
to the 51 neurons in the middle layer, including identical levels of internal noise.
Bottom panel:  a noiseless counter firing twice per $51$ spikes will implement 
a median measurement. Notice that both middle and bottom panels now show {\em two}
spikes per each spike of the middle layer neurons, and one of them (the closest to the
median) is the most reliable. }} 
\end{figure} 

We think that this very simple concept might be of physiological relevance, in
that it implies a trade-off between temporal reliability and the dynamic range
of the signal: when the dynamic range is large, it is possible to afford the
cost of losing part of the charge to achieve reliability, while in a situation
in which the dynamic range is limited, reliability has to be sacrificed in
order to preserve and propagate the signal. This suggests the possibility
of integration pathways which multiplex the signal, as found in retinal adaptation.

\section{Conclusion}

The relevance of the message-dependent nature of noise in spiking elements is,
to our understanding, two-fold. One aspect is its consequences for the use of
Shannon's Information Theory as a framework to measure the information content
present in the output of a neuron. This is the subject of much work recently
\cite{SPIKES}. In this regard, a basic assumption commonly made is that the
noise introduced by the communication channel is independent of the message
being transmitted, which allows the modeling of a neuron as a Gaussian channel
\cite{SHANNON,THEU}. This is indeed desirable given that departures from the
Gaussian channel have proved to be rather cumbersome; for instance, no general
theory for the computation of channel capacity exists \cite{ASH}. Our results
render average measures of information per spike less meaningful than usually
thought, and speak for the necessity of concentrating on particular individual
spikes. This is specially critical when the message consists of only a few
spikes (for instance, in the auditory pathway), or in the event of fast
conductance modulations \cite{FREGNAC}, which can dramatically affect the
reliability. Thus, perhaps the most fundamental consequence of Mainen $\& $
Sejnowski's result \cite{SEJNOWSKI} is the implicit demonstration that cortical
neurons are not always classical Gaussian channels.

A second relevant aspect is the possible anatomical correlate of this
phenomenon. In this regard, we have shown that unless dedicated architecture is
implemented to reduce the multiplication of noise along the processing
pathways, the encoding of behaviorally relevant (not only ``artificial''
stimuli) can be highly degraded. Finally, we demonstrated that it is possible
to design network topologies with arbitrarily large temporal reliability, and
consequently one should expect an evolutionary pressure to implement them in
specific areas of the brain where time accuracy is essential.

Supported in part by the Mathers Foundation (MM), NIH (JMA), Human Frontiers
(LM), Rosita \& Norman Winston Foundation (GAC) and Burroughs Wellcome (MS).
Supercomputer resources are supported by NSF ARI Grants. Discussions with
T. Gardner, S. Ribeiro and R. Crist are appreciated. We also acknowledge the valuable input from D. Reich and B. Knight.


\begin{thebibliography}{99}
\bibitem{simula}In Fig. 1, Eq. 1 was numerically solved  with
parameters: $\sigma=0.01$, $V_{0} = 0.3$; steady input $I=0.8$,
fluctuating input was generated as an Ornstein-Uhlenbeck process with
variance 0.01 and correlation time 0.1. (for easy of presentation $I(t)$
is plot as a ten points running average). In Fig. 3, Eq. 1 parameters
are: noise variance 0.003, $g$=0.1, high input $I$=1.0, low $I$=0.1;
the parameters for Eq. 2-3 are: noise variance 0.003, high input $I=1$,
low $I=0.5$, $\phi$ = 0.08, $a$ = 0.7, and $b$ = 0.8.
\end{thebibliography}
\end{document}